\documentclass[%
 pre,
 amsmath,amssymb,
 reprint,%
]{revtex4-1}

\usepackage{graphicx}
\usepackage{dcolumn}
\usepackage{bm}

\usepackage[utf8]{inputenc}
\usepackage[T1]{fontenc}
\usepackage{mathptmx}
\usepackage{etoolbox}
\usepackage{soul}
\usepackage{xcolor}
\sethlcolor{yellow}

\usepackage{hyperref}

\begin{document}

\preprint{JNRL-xxxxx}

\title{Impact of super-Gaussian electron distributions on plasma K-shell emission}
\author{H. P. Le}
 \email{hle@llnl.gov}
\author{E. V. Marley}%
\author{H. A. Scott}%
\affiliation{ 
Lawrence Livermore National Laboratory, Livermore, California 94550, USA
}%

\date{\today}

\begin{abstract}
Electron distributions in laser-produced plasmas will be driven toward a super-Gaussian distribution due to inverse bremsstrahlung absorption [Langdon, Phys. Rev. Lett.  \textbf{44}, 575 (1980)]. 
Both theoretical and experimental evidence suggest that fundamental plasma properties are altered by the super-Gaussian distribution. 
This paper examines how the super-Gaussian distribution affects the ionization balance and K-shell emission of atomic plasmas, utilizing approximate formulas and detailed collisional-radiative simulations. 
While the impact on plasma ionization is small, K-shell spectra can be significantly modified. Based on these findings, we demonstrate that K-shell spectroscopy can be used to infer super-Gaussian or other similar non-equilibrium electron distributions.
\end{abstract}

\maketitle

\section{Introduction}
\label{sec:intro}
A. B.  Langdon's seminar paper in 1980 predicted that electron distributions in laser-produced plasmas will be driven toward a {{non-Maxwellian}} distribution due to the combined effects of electron oscillation in the high-frequency electric field and electron-ion scattering \cite{langdon_nonlinear_1980}. 
This process leads to a reduction in the inverse bremsstrahlung absorption of the laser light. 
{Mora and Yahi \cite{mora_thermal_1982} and Matte et al. \cite{matte_non-maxwellian_1988} proposed that these laser-heated distributions can be characterized by a super-Gaussian distribution, with the order of the super-Gaussian determined from the numerical solution of the electron kinetic equation.}
Over the last few years, a number of experiments at University of Rochester’s Laboratory for Laser Energetics (LLE) have provided strong evidence supporting the existence of super-Gaussian electron distribution functions in laser-produced plasmas, consistent with {theory} \cite{turnbull_impact_2020,milder_measurements_2021,turnbull_inverse_2023}. 
These experiments utilized Thomson scattering measurements to infer the electron distribution.  

Both theory and experiment indicate that many fundamental plasma properties, such as inverse bremsstrahlung absorption rates, transport coefficients,  and atomic populations are altered because of the super-Gaussian distribution \cite{mora_thermal_1982, turnbull_inverse_2023,sherlock_inverse_2024,bissell_super_2013,shaffer_thermal_2023,Town_fokker_1995,le_influence_2019,milder_evolution_2020}. 
This work examines the effect of the super-Gaussian electron distribution on the {non-local thermodynamic equilibrium (NLTE)} atomic populations and plasma radiative properties, with a particular focus on K-shell emission due to its common use in X-ray spectroscopy. 
{While it is generally accepted that the bulk of the electron distribution can be characterized by a super-Gaussian distribution, there are outstanding questions regarding the high-energy tail--specifically, whether it follows a Maxwellian distribution or a super-Gaussian distribution of different orders. \mbox{\cite{fourkal_electron_2001, milder_measurements_2021,shaffer_thermal_2023}}. Although this study does not directly address the issue of the high-energy tail, as we focus solely on a single super-Gaussian distribution, the understanding gained from this study can aid in the experimental design necessary to explore this question further}. This point will be briefly discussed in Sec.~{\ref{sec:expt}}

Plasma X-ray spectroscopy is a powerful non-intrusive diagnostic technique capable of providing a wealth of information about the composition, density, and temperature of the plasma \cite{hammel_xray_1993,regan_shell_2002,chen_krypton_2017,perez_understanding_2019,Perez_Callejo_demostration_2021,gao_hot_2022}.    K-shell line ratios, such as the those of He$\alpha$ or He$\beta$ to their Li satellites, are often used to determine plasma temperature.  Additionally,  K-shell line widths are commonly used to infer electron density.  The accuracy of the inferred plasma conditions hinges on a physically sound model for simulating X-ray spectra.

Spectroscopic modeling typically involves two main steps: atomic kinetics and spectral calculations. In the atomic kinetics step, the populations of energy levels are determined by considering various atomic processes such as excitation, ionization, and recombination. Once the level populations are known, radiative properties of the plasma, i.e., absorption and emission coefficients, can be calculated for all atomic transitions, including bound-bound, bound-free, and free-free processes. Both atomic kinetics and spectral calculations depend on the electron distribution function.  Non-Maxwellian electron distributions can directly modify the spectra through absorption and emission rates, or indirectly through atomic populations \cite{matte_non-maxwellian_1988,lamoureux_atomic_1993}.

The effects of non-Maxwellian distributions,  particularly those containing a hot electron population, have been a subject of investigation by various researchers. \cite{abdallah_electron_1999,hansen_hot-electron_2002,fournier_influence_2003,hansen_effects_2004}.  Abdallah et al.  examined the effect of energetic electron beams on the spectroscopic measurement in plasma focus experiments \cite{abdallah_electron_1999}. Their model assumed the electron distribution consists of a thermal Maxwellian component plus a Gaussian distribution centered at the energy of the electron beam. Hansen et al. conducted a similar study to examine L-shell spectra of laser-irradiated Krypton cluster \cite{hansen_hot-electron_2002}.  Fournier et al. studied the effect of optical thickness and hot electrons on the Rydberg spectra of laser-produced copper plasmas \cite{fournier_influence_2003}. 
Hansen and Shlyaptseva later generalized their work to systematically investigate the effects of several parameterized non-Maxwellian electron distributions on modeled X-ray spectra \cite{hansen_effects_2004}.  
These studies showed that hot electrons can amplify the intensities of emission lines whose emitting states are populated by collisional excitation processes. 
The situation is reversed for a super-Gaussian distribution, because the tail of the distribution is depleted rather than enhanced.  
Nevertheless, this observation suggests that K-shell spectra can be altered due to the super-Gaussian distribution. 

K-shell line emission takes place when an excited state of a H- or He-like ion undergoes a radiative decay to the ground state. 
The emission intensity of each transition is directly proportional to the density of the excited state. 
In the low-density (coronal) limit, this density is determined by the ratio of the collisional excitation rate to the radiative decay rate between the ground and excited states. 
If the collisional rate is altered by a super-Gaussian distribution \cite{alaterre_ionization_1986,lamoureux_electron_1986}, the resulting line emission will be modified accordingly.  This effect was first discussed by Lamoureux et al.  using a simple model \cite{lamoureux__1989}. 
With increasing plasma density, the atomic kinetics becomes considerably more complicated, requiring a collisional-radiative (CR) modeling approach \cite{ralchenko_modern_2016}. 

In this work, we develop a non-Maxwellian CR model to calculate the population kinetics and K-shell emission of atomic plasmas.  
While the model does not make any approximation to the shape of the electron distribution, the results presented here are limited to super-Gaussian distributions. 
Details of the CR model are discussed in Sec. \ref{sec:cr}. 
An approximate formula is derived for the ratio of collisional rates calculated using super-Gaussian distributions compared to their Maxwellian values. This formula is useful for estimating the effects on K-shell spectra.
CR simulations of Titanium plasmas are shown in Sec. \ref{sec:sim}, where we analyze the difference between super-Gaussian and Maxwellian results. Sec. \ref{sec:expt} builds on these results and discusses the potential for diagnosing super-Gaussian distributions using K-shell spectroscopy.  Finally, a summary is given in Sec. \ref{sec:conclusion}.

\section{Collisional-Radiative Model}
\label{sec:cr}
The CR model for non-Maxwellian {NLTE} plasmas is discussed in this section. 
We only highlight aspects of the CR model which require modification due to the non-Maxwellian distribution. 
The reader is referred to \cite{ralchenko_modern_2016} for an extensive review of the state of the art of CR modeling. 

The distribution of atomic populations $\mathbf{y}$ is determined by the system of rate equations:
\begin{equation}
\label{eq:rateeq}
\frac{d \mathbf{y}}{ dt}= \mathbf{A} \,  \mathbf{y}
\end{equation} 
where the rate matrix $\mathbf{A}$ includes all transition rates (collisional and radiative) between pairs of levels. Numerical solution of Eq. ~(\ref{eq:rateeq}) is computed subject to a constraint of density conservation $\sum_i y_i = N$, where $N$ is the total number density of the plasma.  The free electron density is assumed to follow from the charge-neutrality condition $\sum_i z_i y_i = N_e$.

For many problems, the timescales of atomic kinetics are much faster than the timescale of interest such that the steady-state limit applies, i.e.,  $d \mathbf{y} / dt=0$. In that case, Eq.~(\ref{eq:rateeq}) reduces to a set of algebraic equations, and the atomic populations are completely determined based on the plasma density, electron and photon distributions.  For the results presented here, we first solve Eq.~(\ref{eq:rateeq}) to obtain the atomic populations at a fixed electron density $N_e$,  and then use them to calculate the emission spectra.

For K-shell transitions with large optical depth, the atomic populations can be modified due to trapping of the radiation \cite{Perez_Callejo_demostration_2021}. This effect can be accounted for by the method of escape factors or by coupling the CR model to the radiation transport equation \cite{perez_understanding_2019,london_radiation_2024}. For simplicity, this study assumes that the plasma is optically thin, i.e., zero radiation field. 

When the electron distribution is non-Maxwellian, the macroscopic rate of a transition involving free electrons (collisional excitation/ionization, three-body recombination and radiative recombination) is no longer a function of the plasma temperature.  Instead, it needs to be calculated directly by integrating the microscopic collision rate (cross section × electron velocity) over the electron distribution. For example, the rate for a collisional excitation (or ionization) is:
\begin{equation}
\label{eq:ratex}
R^{\textrm{exc/ion}}= \int f(\varepsilon) \sigma^{\textrm{exc/ion}}  v  \,  d\varepsilon
\end{equation}
where $v = \sqrt{2\varepsilon/m}$.  Numerical evaluation of Eq.~(\ref{eq:ratex}) for a large number of transitions makes non-Maxwellian CR calculations computationally expensive. 
The most expensive transitions are three-body recombination rates since they involve a double integral over the distributions of primary and secondary electrons.
\begin{equation}
\label{eq:rate3br}
R^{\textrm{3br}}= \int \int f(\varepsilon')  f(\varepsilon'') \sigma^{\textrm{3br}}  v' v''  \,  d\varepsilon' \,  d\varepsilon''
\end{equation}
It is worth noting that when the electron distribution is non-Maxwellian, the principle of detailed balance, often used to relate forward and backward rates, must be applied to the differential cross sections for ionization and recombination via the Fowler relation \cite{oxenius_kinetic_1930}. 
Here we use the differential cross section for ionization proposed by Clark, Abdallah and Mann \cite{clark_integral_1991}, which varies from a flat cross section for low total energy of the outgoing electrons to one which increasingly favors one energetic electron rom high total energy.

\begin{center}
\begin{figure*}
\includegraphics[scale=.7]{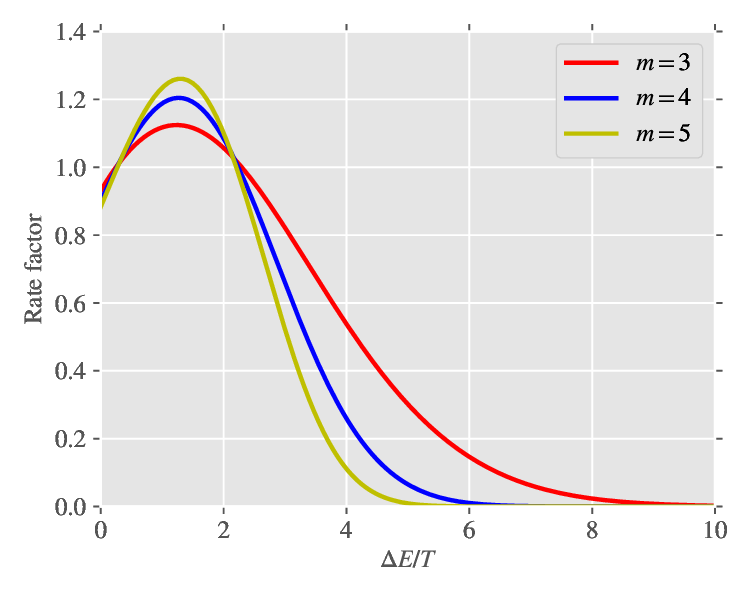}
\includegraphics[scale=.7]{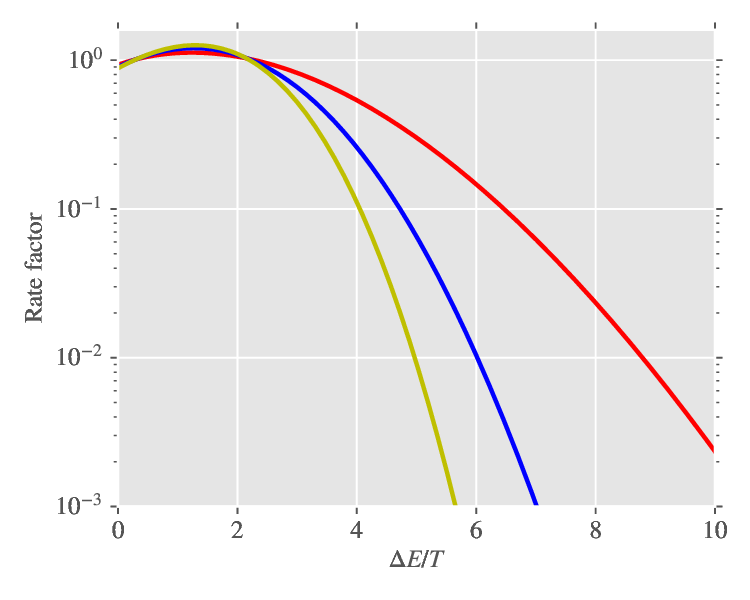}
\caption{Ratios of super-Gaussian excitation/ionization rates (assuming constant collision strength) to their Maxwellian values.}
\label{fig:rate}
\end{figure*}
\end{center}

The super-Gaussian electron energy distribution {due to inverse bremsstrahlung absorption is assumed to take the form \cite{mora_thermal_1982,matte_non-maxwellian_1988}}:
\begin{equation}
\label{eq:fsg}
f( \varepsilon ) 
 \propto \exp \left[ -\left( a_m \varepsilon/T \right)^{m/2}  \right]
\end{equation}
where $a_m=2 \Gamma (5/m) / 3 \Gamma (3/m)$ and $\Gamma(s)= \int_0^\infty t^{s-1} e^{-t} \, dt$  is the Gamma function. The order of the super-Gaussian $m$ is related to a dimensionless parameter $\alpha$, defined as the ratio of the electron-ion energy transfer rate to the electron-electron thermalization rate:
\begin{equation}
\label{eq:alpha}
\alpha = Z^* v_{\textrm{osc}}^2/v_t^2
 \end{equation}
 where $v_{\textrm{osc}}$ is the oscillation velocity of electrons due to the laser electric field and $v_t = \sqrt{T/m}$ is the thermal velocity.  The relationship between $\alpha$ and $m$ was obtained from numerical simulations and fitted according to the following formula \cite{matte_non-maxwellian_1988}:
\begin{equation}
\label{eq:matte}
m(\alpha) = 2 + 3/(1 + 1.66/\alpha^{0.724})
\end{equation}

The numerical fit (\ref{eq:matte}) has two opposite limits: $m=2$ for $\alpha \ll 1$ (strong thermalization) and $m=5$ for $\alpha \gg 1$ (weak thermalization).  In the strong thermalization limit, Eq. ~(\ref{eq:fsg}) yields the Maxwellian distribution.  For $m>2$, the super-Gaussian distribution (\ref{eq:fsg}) features a depleted high energy tail due to the fast drop-off at high energy.  

To illustrate how a super-Gaussian distribution (\ref{eq:fsg}) modifies the transition rate shown in Eq.~(\ref{eq:ratex}), let us consider a collisional excitation (or ionization) where the collision strength $\varOmega \equiv \sigma \varepsilon$ is assumed to be constant. 
In this case, the ratio of the collisional rate calculated with a super-Gaussian distribution of exponent $m$ over its Maxwellian value is:

\begin{equation}
\label{eq:sgfac}
\frac{R_{m}}{R_{m=2}} = \frac{\sqrt{\pi}}{2} a_m^{1/2} e^{\Delta E/T} \frac{\Gamma \left[ 2/m, (a_m \Delta E /T)^{m/2}\right]}{\Gamma \left( 3/m \right)}
\end{equation}
where $\Gamma (s,x) = \int_x^\infty t^{s-1} e^{-t} \, dt$ is the incomplete Gamma function. 
Eq.~(\ref{eq:sgfac}) is plotted in Figure~\ref{fig:rate} for a range of transition energies $\Delta E/T$ and values of $m$.  Here the temperature for a non-Maxwellian distribution is defined such that $\int_0^\infty f(\varepsilon) \, \varepsilon \, d\varepsilon = \frac{3}{2} N_e T$.  The result of Eq.~(\ref{eq:sgfac}) for $\Delta E/T=0$ is directly applicable to collisional deexcitation rates where super-Gaussian distributions result in approximately $10\%$ reduction.  For three-body recombination rates shown in Eq.~(\ref{eq:rate3br}), Eq.~(\ref{eq:sgfac}) can be applied twice, one for each electron, resulting in approximately $20\%$ reduction.

For a transition with a finite energy threshold,  super-Gaussian distributions can result in either a lower or higher rate than the Maxwellian distribution, depending on the value of $\Delta E/T$.  
For $\Delta E/T \approx 1.3$, super-Gaussian rates are 10-30\% higher than their Maxwellian values.  
For larger values of $\Delta E/T$ ($ > 3$), super-Gaussian rates are significantly lower than their Maxwellian values. 
This is because transition rates with large energy threshold (compared to the average energy of the electrons) are very sensitive to the tail of the distribution,  where super-Gaussian and Maxwellian distributions exhibit the largest difference.

Numerical experiments indicate that Eq.~(\ref{eq:sgfac}) provides a {reasonably accurate} approximation to numerically calculated super-Gaussian rates for a wide range of conditions and cross section fits.  These ratios can be used to modify standard CR models to account for super-Gaussian and possibly other non-Maxwellian distributions. 
Implementing these modifications in existing CR codes should be straightforward and does not incur any significant increase in computational time. However, the results shown in the present study are obtained through direct numerical integration of the electron distribution, and therefore, involve no approximation.

\section{Numerical simulations}
\label{sec:sim}

\begin{center}
\begin{figure*}
\includegraphics[scale=.7]{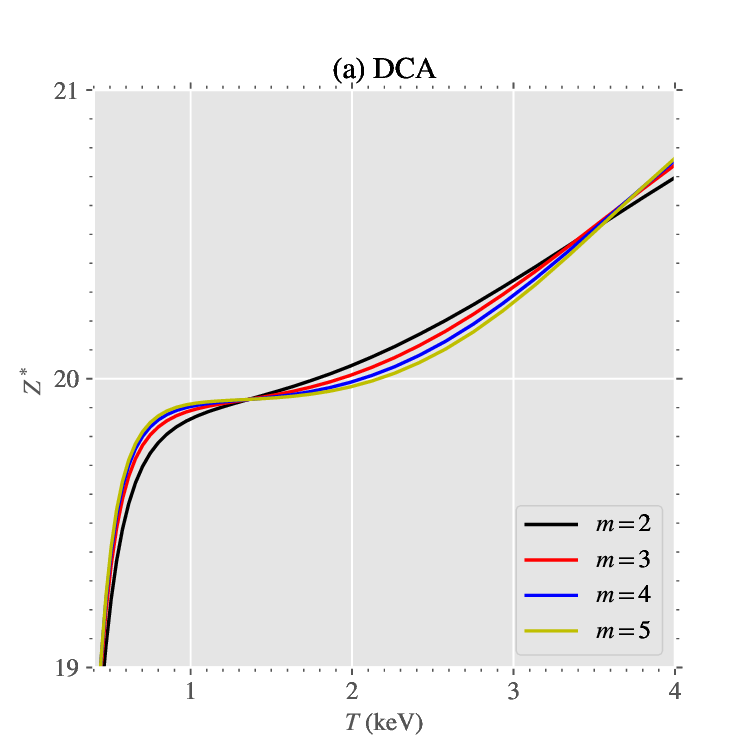}
\includegraphics[scale=.7]{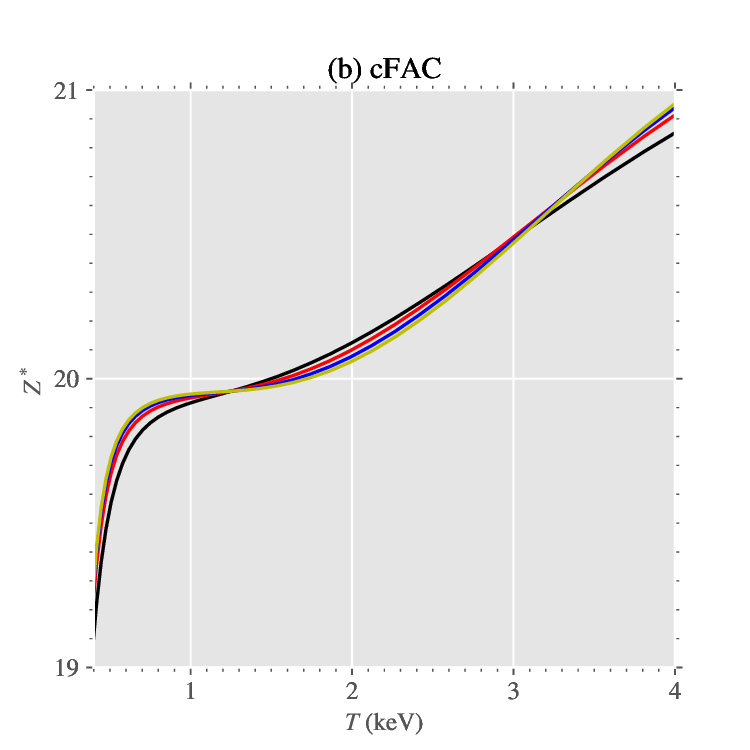}
\caption{Average ionization $Z^*$ of a Ti plasma at $N_e = 10^{21}$ cm$^{-3}$ as a function of temperature using (a) DCA and (b) cFAC atomic models.}
\label{fig:zbar}
\end{figure*}
\end{center}

\begin{center}
\begin{figure*}
\includegraphics[scale=.7]{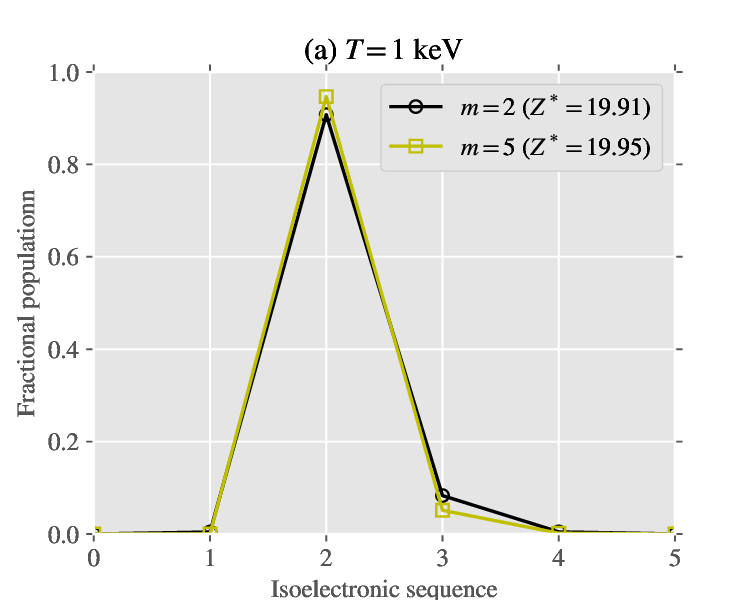}
\includegraphics[scale=.7]{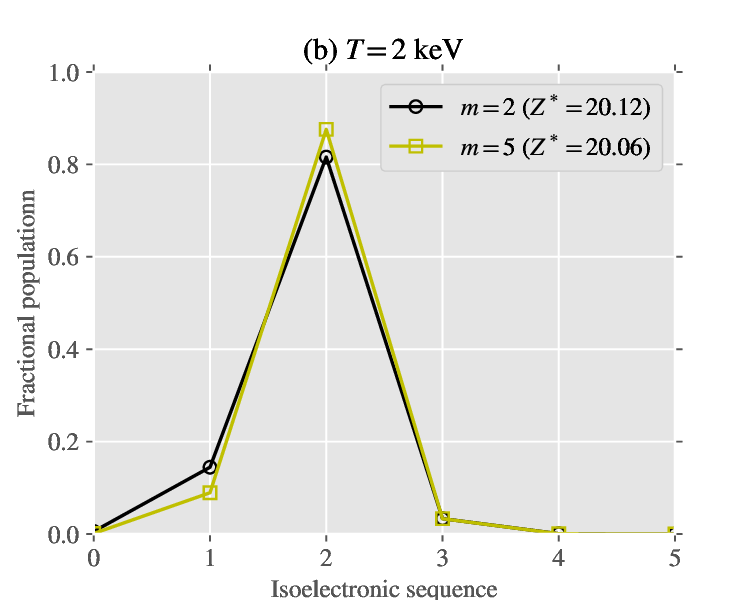}
\caption{Charge state distribution of Ti plasma at $N_e = 10^{21}$ cm$^{-3}$ and temperatures of (a) 1 keV and (b) 2 keV calculated using the cFAC atomic model.}
\label{fig:csd}
\end{figure*}
\end{center}

The non-Maxwellian CR model discussed in the previous section is implemented in the atomic kinetics and radiation transport code Cretin \cite{scott_cretin_2001}.  
In this section, we show Cretin simulations of a titanium (Ti) plasma, and examine the effects of super-Gaussian distributions on the charge state balance and K-shell emission. Similar simulations using other elements (results not shown here) confirm that the effects remain consistent. We will show that the degree to which the super-Gaussian distribution affects the spectra depends sensitively on the transition energy--specifically the value of $\Delta E/T$. This sensitivity is exploited in Sec.~\ref{sec:expt} to aid in the design of K-shell spectroscopy experiments to {infer the super-Gaussian distribution}.

The atomic kinetics calculation requires input data for energy levels and transition rates, which can be constructed from different underlying atomic physics models. 
To demonstrate that the impact of super-Gaussian distributions is not unique to a particular atomic model choice, we consider two sets of atomic physics models. 
The first set consists of the so-called DCA models, which are widely used in radiation-hydrodynamics simulation. The use of "DCA" does not refer to the detailed configuration accounting description of atomic structure. 
These models are constructed with the screened hydrogenic formalism for energy levels and cross sections with energy levels described by principal quantum numbers.  
These models include single and double excitations from the valence shell up to principal quantum number of 10, along with a few excitations from inner shells. Further details about the DCA models can be found in \cite{scott_advances_2010}. Due to the highly averaged nature of the atomic structure, these models are not suitable for producing detailed K-shell spectra. 

The second set of models is based on cFAC, the C-variant of the fully relativistic Flexible Atomic Code (FAC) \cite{gu_flexible_2008,cfac}. 
The original version of FAC operates in two distinct modes: fine structure levels or relativistic configuration averages. 
While the fine structure levels are required for accurate K-shell spectra, producing atomic models with broad coverage in atomic energy phase space leads to a significant number of levels and transitions. 
cFAC features a hybrid fine-structure and configuration average mode, which allows us to construct atomic models that are computationally tractable, have the same coverage as DCA models, but with much higher spectral accuracy.

{
In this study, we focus on plasma conditions where the NLTE effect is strong (low density and high temperature), and the electron distribution is laser-driven toward a super-Gaussian profile. This occurs below the laser's critical density, e.g., $N_{\textrm{cr}} = 9 \times 10^{21}$ cm$^{-3}$ for a typical laboratory 3$\omega$ laser. Previous studies have shown that K-shell spectra are not sensitive to electron density within this range \cite{marley_using_2018, bishel_ionization_2023}. For this reason, we present simulation results for a representative condition of $N_e = 10^{21}$ cm$^{-3}$. Additionally, we show Ti K-shell spectra due to their relevance in previous laser-produced plasma experiments investigating NLTE plasmas \cite{marley_using_2018}. The impact of the super-Gaussian distribution on plasma K-shell emission is similar for other elements, although it occurs at different temperatures due to variations in K-shell binding energies.
}

Figure~\ref{fig:zbar} shows the average ionization of a Ti plasma at constant electron density $10^{21}$ cm$^{-3}$ as a function of temperature for different super-Gaussian distributions ($m=2$ corresponds to a Maxwellian) using (a) DCA and (b) cFAC atomic models. 
Results from DCA and cFAC agree reasonably well, particularly near the closed shell configuration, where the He-like emission is the highest. 
In this temperature range, the effect on the ionization balance is small but consistent between the two atomic models.  
Specifically, super-Gaussian distributions result in a slightly higher ionization at temperatures approaching He-Like ($Z^* = 20$),  and lower ionization past He-like. The same pattern repeats as the ionization approaches H-like ($Z^*=21$). 

These results show that super-Gaussian distributions have a minor effect on the average ionization of the plasma. However, this effect is physical and can be understood by examining Figure~\ref{fig:rate} or applying Eq. ~(\ref{eq:sgfac}) to the collisional ionization rates of Li-like and He-like ions. As will be discussed later, the effect on K-shell emission is qualitatively similar but more significant. Therefore, it is important to explain the cause of the difference in average ionization due to super-Gaussian distributions.

\begin{center}
\begin{figure*}
\includegraphics[scale=.58]{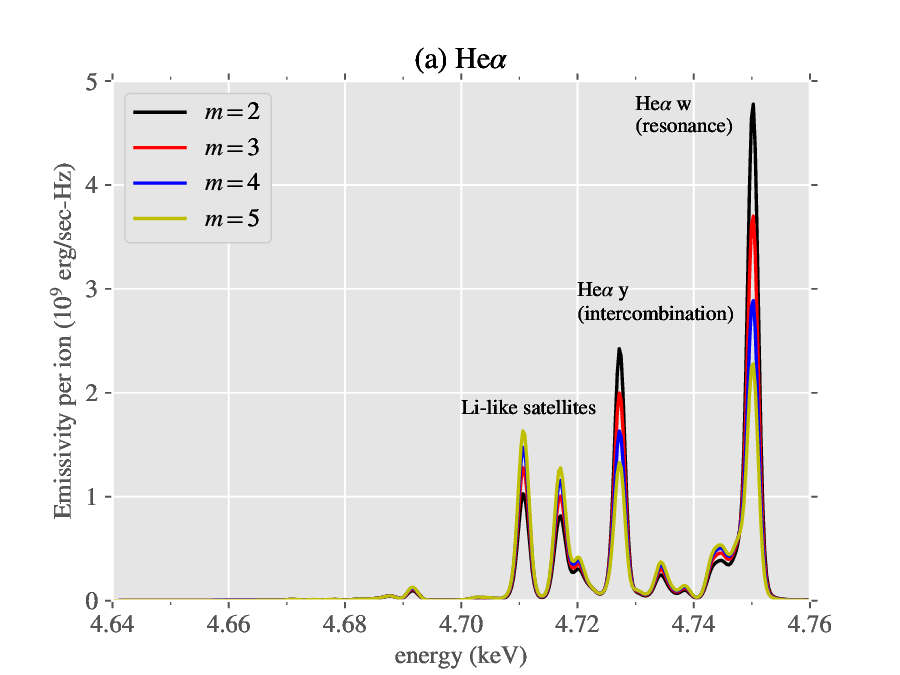}
\includegraphics[scale=.58]{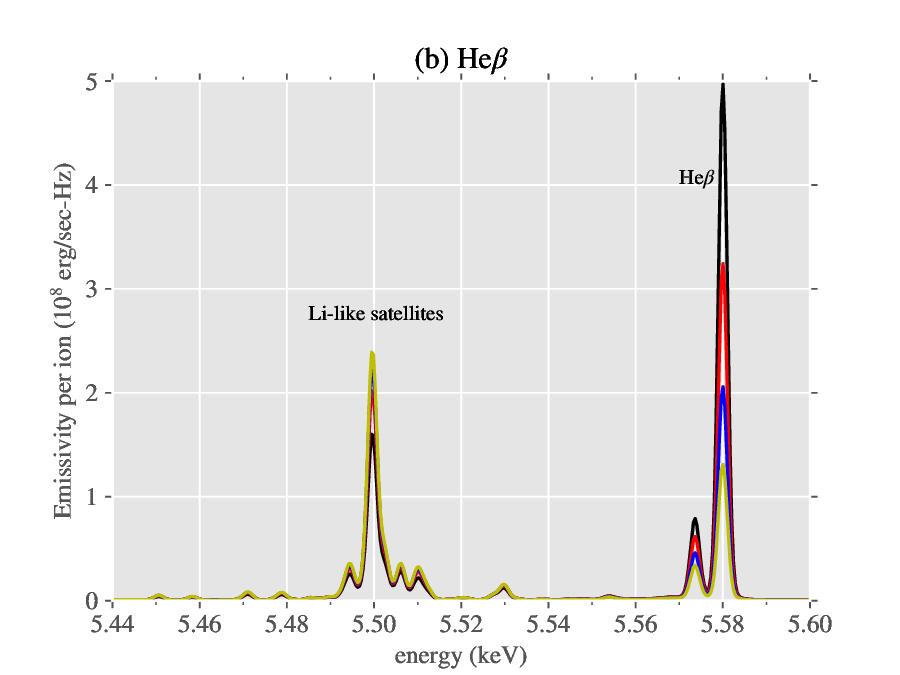}
\caption{He K-shell emission spectra for Ti plasma at $N_e = 10^{21}$ cm$^{-3}$ and $T = 1.5$ keV: (a) He$\alpha$ and (b) He$\beta$}
\label{fig:he1500}
\end{figure*}
\end{center}

\begin{center}
\begin{figure}
\includegraphics[scale=.7]{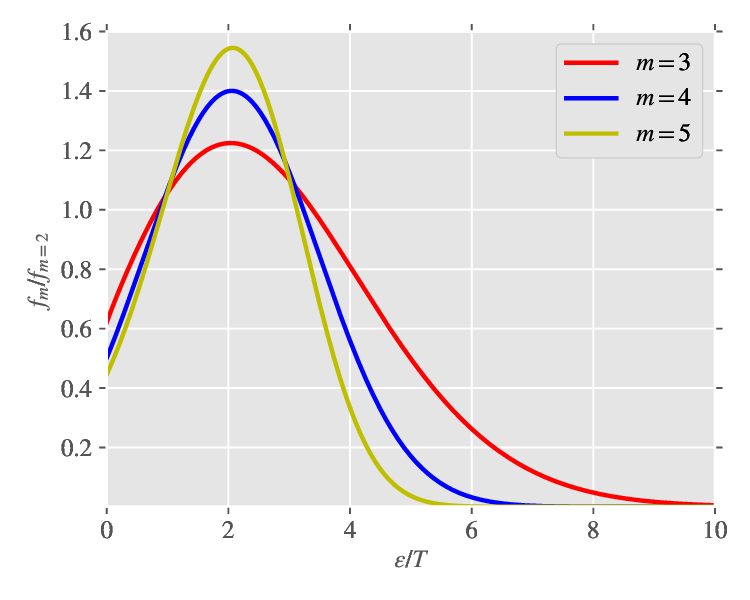}
\caption{Ratios of radiative capture rates for super-Gaussian distributions to their Maxwellian values.}
\label{fig:rate3}
\end{figure}
\end{center}

\begin{center}
\begin{figure*}
\includegraphics[scale=.58]{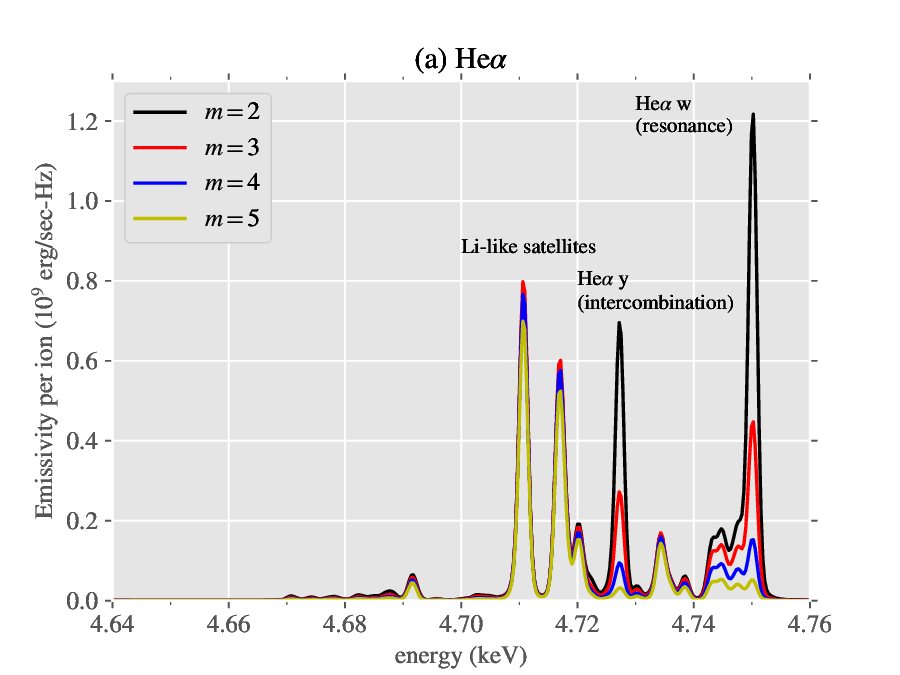}
\includegraphics[scale=.58]{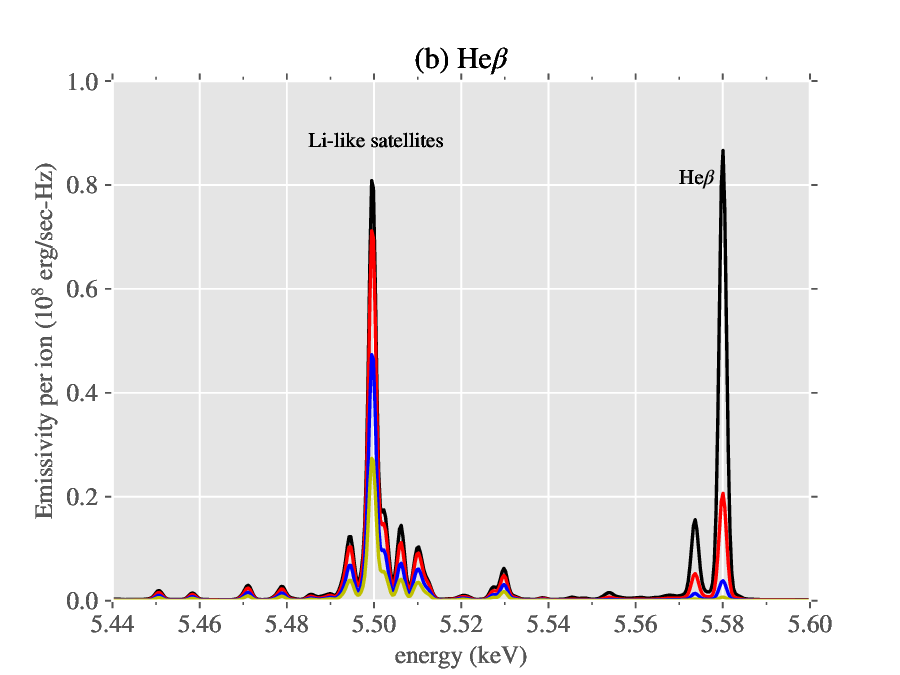}
\caption{He K-shell emission spectra for Ti plasma at $N_e = 10^{21}$ cm$^{-3}$ and $T = 1$ keV: (a) He$\alpha$ and (b) He$\beta$}
\label{fig:he1000}
\end{figure*}
\end{center}

Figure~\ref{fig:csd} shows the charge state distributions at $T=1$ and 2 keV using the cFAC atomic model. 
Results from the DCA model are similar and not shown here. 
Super-Gaussian distributions have the opposite effect on average ionization, i.e., higher $Z^*$ at 1 keV and lower at 2 keV, compared to Maxwellian results. 
This can be explained by applying Eq. ~(\ref{eq:sgfac}) to the collisional ionization of Li-like and He-like ions.
The ionization energies of Li-like and He-like are approximately 1.4 and 6.2 keV. 
At $T=1$ keV, the super-Gaussian distribution leads to increased ionization rates of Li-like ions and decreased ionization rates of He-like ions. 
The recombination rates are also altered, but their impact, approximately a 10\% decrease, is smaller relative to collisional ionization rates.  
Decreased collisional ionization rates of He-like ions at this temperature are less significant because their magnitudes are very small ($R^{\textrm{ion}} \propto e^{-\Delta E/T}$), while increased collisional ionization rates of Li-like ions ultimately lead to a slightly lower Li-like and a higher He-like population at 1 keV (Figure~\ref{fig:csd}a).
At 2 keV, the He-like ionization rates are larger ($\Delta E / T \approx 3$), so H-like ions are significantly populated. Here the results of Eq. ~(\ref{eq:sgfac}) indicate lower collisional ionization rates of He-like ions for a super-Gaussian distribution, which leads to a preferentially less ionized charge state balance, i.e., lower H-like and higher He-like populations (Figure~\ref{fig:csd}b). Similar simulations for other elements typically used in K-shell diagnostics verifies that the same effect is consistently observed among these ions.  

The effect of super-Gaussian distributions on modeled K-shell spectra is shown next. Here, we focus on the 1-2.5 keV temperature range where the He-like emission is the largest.  {For simplicity, we assume that the plasma is optically thin, so the radiation intensity is directly proportional to the emission coefficients. For K-shell transitions with large optical depth, radiation trapping can modify the atomic populations, resulting in attenuated K-shell line intensity. Although a complete treatment requires a self-consistent solution of the radiation transport and atomic kinetics equations, a simple estimate based on the escape factor formalism shows that the sensitivity of the K-shell spectra to the super-Gaussian distribution is independent of the optical depth effect. The comparison of the charge state distribution indicates that the ground state populations are not affected by the super-Gaussian distribution; therefore, the optical depths of the K-shell transitions, which are proportional to the density of the ground state,  are not sensitive to the super-Gaussian distribution. As a result, the escape factor, as well as the attenuation factor--both of which depend strongly on the optical depth--will not be altered by the super-Gaussian distribution.}

Figure~\ref{fig:he1500} shows Ti K-shell emissions from the He$\alpha$ and He$\beta$ complexes at $N_e = 10^{21}$ cm$^{-3}$ and $T = 1.5$ keV calculated using the cFAC atomic model. 
The emission coefficients in these plots are normalized by the ion density, so the differences in the spectra are solely due to the effect of super-Gaussian distributions on the ionization balance and transition rates.  The results demonstrate a high degree of sensitivity of both He$\alpha$ and He$\beta$ complexes to the electron distribution. The calculations using the DCA atomic model show the same effect, albeit with a less resolved spectrum due to the average nature of the atomic model.

\begin{center}
\begin{figure*}
\includegraphics[scale=.65]{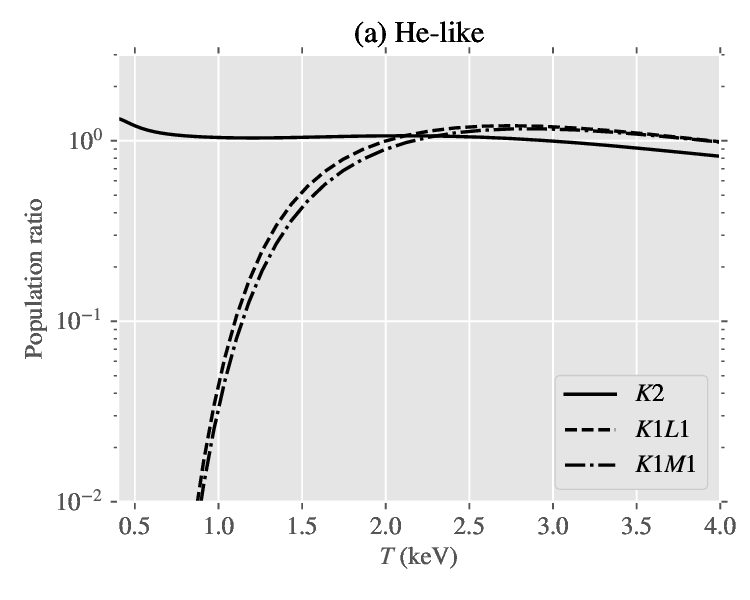}
\includegraphics[scale=.65]{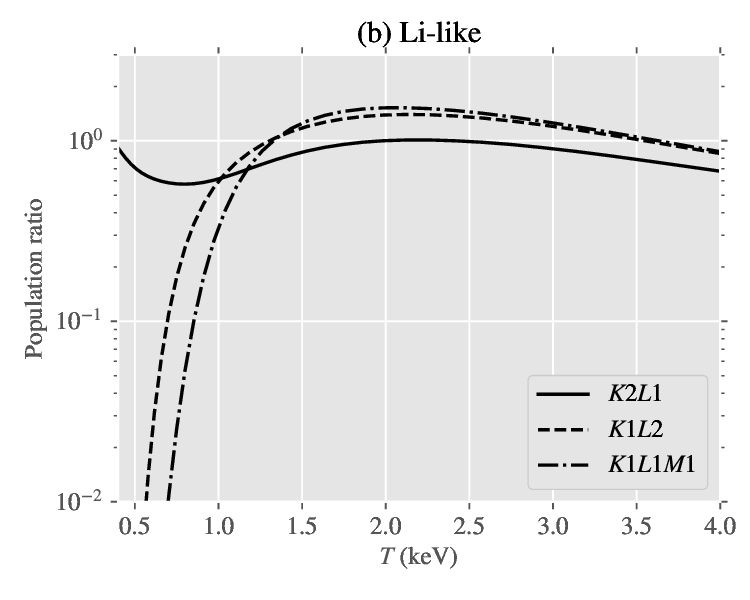}
\caption{{Ratio of super-Gaussian ($m=5$) to Maxwellian excited state populations as a function of temperature for (a) He-like and (b) Li-like Ti ions at $N_e = 10^{21}$ cm$^{-3}$. The K\#L\#M\# notation defines a super-configuration based on the occupancy of the K, L and M shells. Each super-configuration consists of a number of fine levels.The transitions K1L1 $\rightarrow$ K2 and K1M1 $\rightarrow$ K2 denote the main He$\alpha$ and He$\beta$ lines, while K1L2 $\rightarrow$ K2L1 and K1M1L1 $\rightarrow$ K2L1 denote the strongest component of the Li-like satellites with a $n=2$ spectator.}}
\label{fig:pops}
\end{figure*}
\end{center}

\begin{center}
\begin{figure*}
\includegraphics[scale=.58]{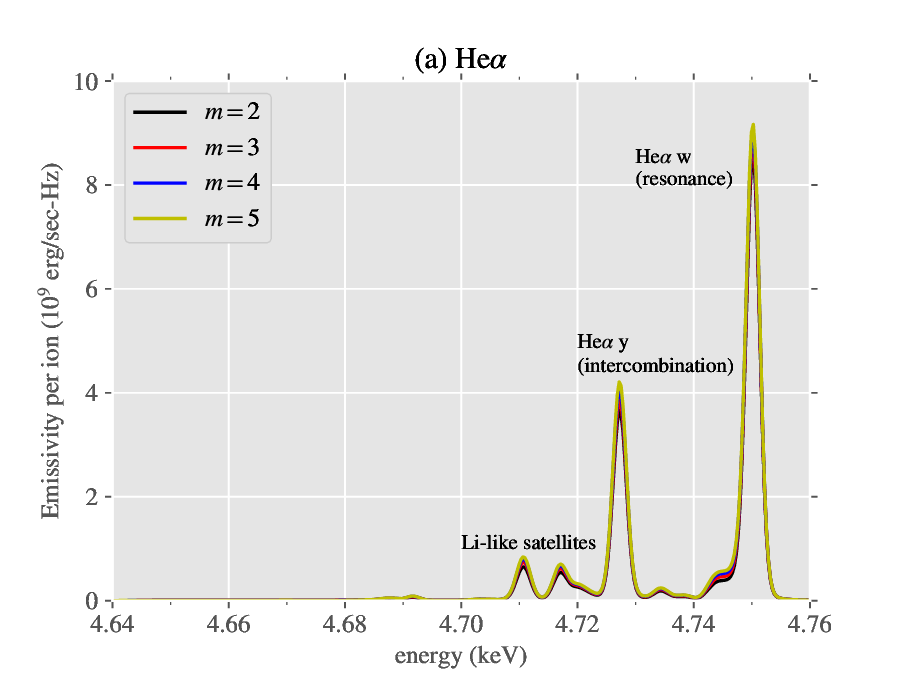}
\includegraphics[scale=.58]{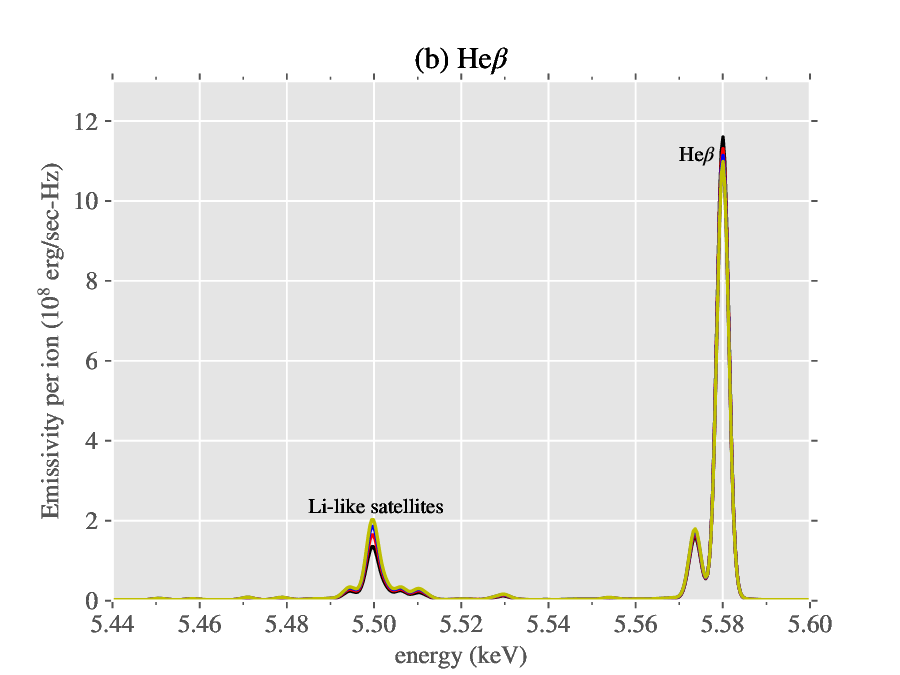}
\caption{He K-shell emission spectra for Ti plasma at $N_e = 10^{21}$ cm$^{-3}$ and $T = 2.5$ keV: (a) He$\alpha$ and (b) He$\beta$}
\label{fig:he2500}
\end{figure*}
\end{center}

\begin{center}
\begin{figure}
\includegraphics[scale=.7]{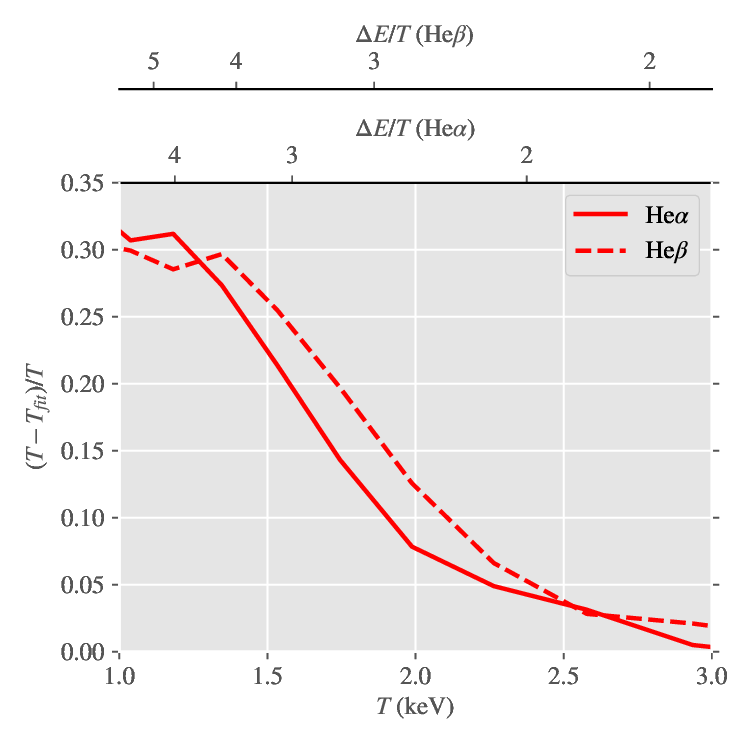}
\caption{Temperature error from fitting individual Ti K-shell spectra (He$\alpha$ and He$\beta$) of super-Gaussian electron distribution ($m=3$) using Maxwellian approximation. The secondary axes at the top show values of $\Delta E/T$ calculated using transition energies of the resonant He$\alpha$ and He$\beta$ lines.}
\label{fig:tfiterror}
\end{figure}
\end{center}

At $T=1.5$ keV, super-Gaussian distributions decrease the emission of both He$\alpha$ and He$\beta$ lines while increasing the emission of their associated satellites. 
This indicates that when the electron distribution is super-Gaussian, assuming a Maxwellian distribution will underestimate the temperature, as a higher ratio of satellites to the main line suggests a colder plasma.
The reduced emission of He$\alpha$ and He$\beta$ is due to smaller collisional excitation rates responsible for populating the upper states. 
The transition energies of the resonant He$\alpha$ and He$\beta$ lines are 4.75 and 5.58 keV, respectively, so super-Gaussian rates can be reduced by up to 50-60\% (see Figure~\ref{fig:rate}). 
In contrast, the increased satellite emission indicates a higher population of the upper states. These states are populated by electron capture from He-like ions (dielectronic recombination satellite) or direct inner-shell excitation from Li-like ground state (inner-shell satellite);
however, at this density the inner-shell satellites are weaker than the dielectronic recombination ones.  
The dielectronic recombination is a compound process consisting of an electron capture followed by a radiative stabilization.  The electron capture process is mainly responsible for populating the upper states of the dielectronic recombination satellites. 
The ratio of the super-Gaussian electron capture rates to the Maxwellian value is equal to the ratio of the distributions evaluated at the transition energy:
\begin{equation}
\label{eq:frc}
\frac{f_m}{f_{m=2}} = \frac{\sqrt{\pi} }{4 } \frac{m a_m^{3/2}}{\Gamma(3/m)} \frac{\exp \left[ -\left( a_m \varepsilon/T \right)^{m/2}  \right]}{\exp \left( -\varepsilon/T \right)}
\end{equation}
Figure \ref{fig:rate3} shows the result of Eq.~(\ref{eq:frc}) for different values of $m$. 
While the ratios appear qualitatively similar to those derived from Eq.~(\ref{eq:fsg}), the transition energies at which they intersect the ratio of 1 are different.
Note that the transition energy of the electron capture process that populates the upper state of a satellite line is not the same as the energy at which the line emission occurs; they differ by the ionization energy of the lower state.  Using the results from Figure~\ref{fig:rate3} for $T=1.5$ keV, the super-Gaussian distribution leads to increased radiative capture rates,  resulting in the enhanced emission of the dielectronic recombination satellites. 

At $T=1$ keV, the values of $\Delta E / T$ for both the main lines and electron capture processes associated with the satellites exceed 3, resulting in a reduction in both collisional excitation and radiative capture rates. This suggests that both the main lines and their satellites are suppressed under super-Gaussian distributions, as confirmed by the simulated emission spectra (see Figure \ref{fig:he1000}).
The suppression of the main lines is more pronounced due to a greater reduction in collisional excitation rates.

The effects of super-Gaussian distributions so far can be explained by approximate formula of the rates, but they can also be determined by directly examining the atomic populations. {Figure \ref{fig:pops} shows the ratio of super-Gaussian ($m=5$) to Maxwellian excited state populations of He- and Li-like ions, which are mainly responsible for He$\alpha$ and He$\beta$ emissions}. For clarity, the states shown here are grouped into super-configurations based on the occupation of the K, L and M shells, and the density of each super-configuration is summed over all fine levels within that configuration. In particular, K1L1 $\rightarrow$ K2 and K1M1 $\rightarrow$ K2 denote the main He$\alpha$ and He$\beta$ lines, while K1L2 $\rightarrow$ K2L1 and K1M1L1 $\rightarrow$ K2L1 denote the strongest component of the Li-like satellites with a $n=2$ spectator. {It is evident that while the ground state populations are not significantly influenced by the super-Gaussian distribution, the excited states--particularly those responsible for the main He$\alpha$ and He$\beta$ lines--are strongly affected. This observation aligns with predictions from the approximate formula and the simulated spectra.}  Figure \ref{fig:pops} indicates that the differences between super-Gaussian and Maxwellian spectra diminish significantly at higher temperature (smaller $\Delta E /T$), as confirmed by the simulated spectra at $T=2.5$ keV (see Figure~\ref{fig:he2500}).

It is now apparent that using K-shell spectra to infer temperature without properly accounting for non-Maxwellian distributions can lead to significant errors. 
The magnitude of the error specifically depends on the value of $\Delta E/T$, but it can also be affected by the spectral fitting procedure as well as the deviation from the Maxwellian distribution.  
Figures \ref{fig:he1500} and \ref{fig:he1000} indicate that the errors are largest for $m=5$; however, the laser intensity required to achieve this condition is quite high. 
Typical laboratory plasma conditions are often in the range of $m=2-3$.  
Figure \ref{fig:tfiterror} shows temperature errors incurred when fitting the He$\alpha$ and He$\beta$ spectra calculated using super-Gaussian distributions ($m=3$) at $T=1-3$ keV and $N_e = 10^{21}$ cm$^{-3}$, while assuming Maxwellian distribution.
Here, positive errors mean that the inferred temperature is lower than the actual value when the electron distribution is super-Gaussian. 
The errors can be as large as 30\% at low temperature (high $\Delta E/T$); at higher temperature (low $\Delta E /T $) they are considerably reduced as expected from the analysis presented in this section.

\begin{center}
\begin{figure*}
\includegraphics[scale=.7]{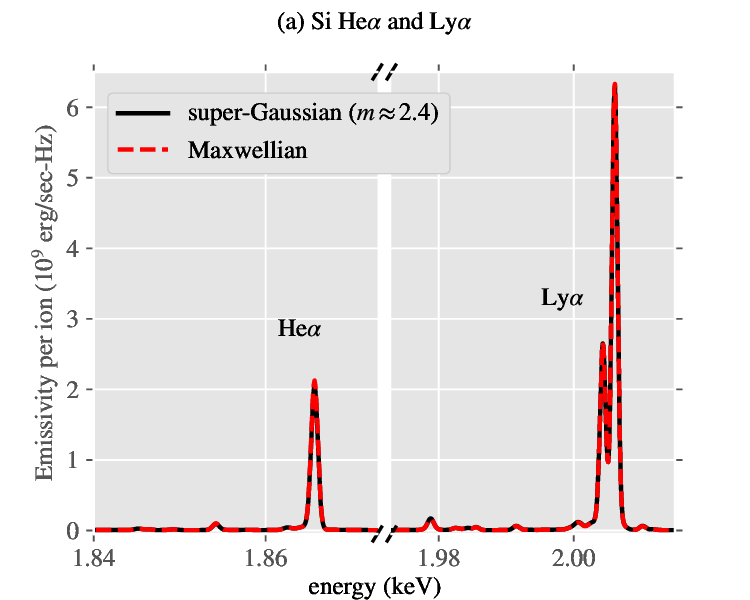}
\includegraphics[scale=.7]{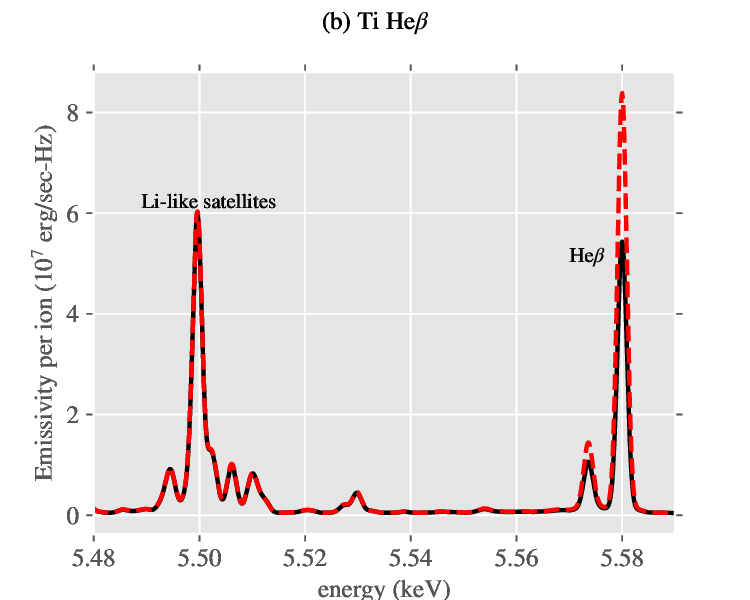}
\caption{(a) Si (He$\alpha$ and Ly$\alpha$) and (b) Ti (He$\beta$) K-shell spectra of a Si-Ti plasma (1:1 mixture) at $N_e = 10^{21}$ cm$^{-3}$ and $T=1.1$ keV. The super-Gaussian results use a value of $m \approx 2.4$ obtained from Eq.~({\ref{eq:matte}}) assuming a 3$\omega$ laser intensity of $10^{15}$ W/cm$^{2}$.}
\label{fig:expt}
\end{figure*}
\end{center}

\section{Diagnosing super-Gaussian distributions using K-shell spectroscopy}
\label{sec:expt}

The results from Figure \ref{fig:tfiterror} indicate the feasibility of { inferring a super-Gaussian electron distribution} using X-ray spectroscopy.  
The difference between the Maxwellian and super-Gaussian K-shell spectra at a given condition depends strongly on the transition energies.
This suggests that by doping the plasma with K-shell tracers of different atomic number, we can effectively probe different part of the energy distribution.  
For a super-Gaussian distribution,  two K-shell tracers--one with $\Delta E/T \approx 2$ and another with $\Delta E/T > 3$--can be used to infer both the temperature and super-Gaussian exponent.

Let us consider an example plasma with $N_e=10^{21}$ cm$^{-3}$ and $T=1.1$ keV. 
Figure~\ref{fig:he1000} shows that both Ti He$\alpha$ and He$\beta$ complexes are sensitive to the super-Gaussian distribution, making it a suitable tracer for inferring the exponent $m$. 
Figure \ref{fig:rate} indicates that for $\Delta E/T \approx 2$, collisional excitation rates are not significantly altered, resulting in similar intensities of these K-shell lines between Maxwellian and super-Gaussian distributions. 
This observation suggests that Silicon (Si) could be used as a low Z tracer.  
At this temperature, Si can be ionized to H-like, making Ly$\alpha$ emission significant and relevant for the analysis. 
The transition energy of Ly$\alpha$ is close to that of He$\alpha$, so the effects of the super-Gaussian distribution are comparable for a given temperature. 
Figure \ref{fig:expt} shows the emission spectra of Si K-shell (He$\alpha$ and Ly$\alpha$) and Ti He$\beta$ for Maxwellian and super-Gaussian distributions. 
The super-Gaussian exponent $m$ is determined self-consistently from Eq.~({\ref{eq:matte}}) assuming a typical 3$\omega$ laser intensity of $10^{15}$ W/cm$^{2}$ and a mixture ratio of 1:1. 
Figure~\ref{fig:expt} demonstrates that by simultaneously fitting the Si and He K-shell spectra, we can infer a super-Gaussian distribution characterized by a plasma temperature $T$ and exponent $m$.

As mentioned earlier, an outstanding question remains regarding whether the high-energy tail of the distribution, in the presence of laser absorption and other transport phenomena, is Maxwellian or super-Gaussian \mbox{\cite{fourkal_electron_2001,milder_measurements_2021,shaffer_thermal_2023}}.
This question can be addressed by introducing an additional K-shell tracer with a higher atomic number specifically designed to probe the high-energy tail.
Using the same plasma conditions in Figure~{\ref{fig:expt}} as an example, an iron tracer can be employed to diagnose the energy distribution in the range of 7-8 keV.

Multiple existing basic-science laboratory platforms could be used to test this idea. 
The laser experiments at LLE mentioned in Sec.~{\ref{sec:intro}} \cite{turnbull_impact_2020,milder_evolution_2020,milder_measurements_2021,turnbull_inverse_2023}, using Thomson scattering analysis, confirmed the presence of super-Gaussian distributions, so can naturally be extended to perform X-ray spectroscopy. 
This experimental configuration is also simple (laser heating a gas jet), and the plasma conditions are well-characterized. 
However, the requirement of multiple K-shell tracers may lead to additional complications.  
Buried layer platforms are also a very attractive candidate for creating plasma conditions where {laser-heated electron distributions can be formed and tested} \cite{ marley_using_2018,Perez_Callejo_demostration_2021,bishel_ionization_2023}. 
These platforms are equipped with multiple spectroscopic diagnostics, and prior experiments have already used mixture of elements to measure different X-ray spectra simultaneously,  such as K- and L-shell.

\section{Conclusion}
\label{sec:conclusion}
In this paper, we have shown how super-Gaussian electron distributions, typically observed in laser-produced plasmas as a result of inverse bremsstrahlung absorption, can modify the charge state balance and K-shell emission spectra of atomic plasmas, using both approximate formula for the atomic transition rates and detailed CR simulations. 
The origin of the difference stems from the modification of transition rates, particularly those that involve free electrons, due to super-Gaussian distributions. 
The approximate formula for super-Gaussian rates derived in this work can be used to estimate the impact of those distributions on the ionization balance and K-shell emission, as confirmed by comparison to detailed CR simulations. 

The present work indicates that under plasma conditions with peaked K-shell emissions, the modification of the ionization balance due super-Gaussian distributions is small; however, K-shell emission spectra can be altered significantly. 
In particular, K-shell transitions at large energy ($\Delta E/T > 3$), are affected the most, because their upper state populations are strongly modified by the super-Gaussian distribution.

The modification of K-shell spectra can be used to design laboratory experiments to infer {laser-heated super-Gaussian electron distributions}.  
We demonstrated that by using two different K-shell tracers in a single plasma, we can infer a super-Gaussian distribution from the measured X-ray spectra. 
This technique can potentially be generalized to infer other non-equilibrium electron distributions. 
For example, non-equilibrium electron distributions with a depleted tail, similar to a super-Gaussian distribution, are commonly observed in the hot region of non-uniform plasmas with strong temperature gradients, where the thermal transport becomes non-local \cite{bell_elecron_1981,matte_electron_1982,albritton_nonlocal_1986}.  
Such distributions can be inferred using the same technique discussed here to identify the onset of non-local thermal transport.

\begin{acknowledgments}
This work was performed under the auspices of the U.S. Department of Energy by Lawrence Livermore National Laboratory under Contract DE-AC52-07NA27344. LLNL-JRNL-2000555

This document was prepared as an account of work sponsored by an agency of the United States government. Neither the United States government nor Lawrence Livermore National Security, LLC, nor any of their employees makes any warranty, expressed or implied, or assumes any legal liability or responsibility for the accuracy, completeness, or usefulness of any information, apparatus, product, or process disclosed, or represents that its use would not infringe privately owned rights. Reference herein to any specific commercial product, process, or service by trade name, trademark, manufacturer, or otherwise does not necessarily constitute or imply its endorsement, recommendation, or favoring by the United States government or Lawrence Livermore National Security, LLC. The views and opinions of authors expressed herein do not necessarily state or reflect those of the United States government or Lawrence Livermore National Security, LLC, and shall not be used for advertising or product endorsement purposes.
\end{acknowledgments}

\nocite{*}
\bibliography{zotero}

\end{document}